%% file: main.tex
\documentclass[runningheads]{llncs}
\usepackage{adjustbox}
\usepackage{booktabs}
\usepackage{tabu}
\usepackage{multirow}
\usepackage{xcolor}
\usepackage{pifont}
\usepackage{amssymb}
\usepackage[outline]{contour}
\usepackage{arydshln}
\usepackage{caption, subcaption}

\usepackage[hidelinks]{hyperref}
\usepackage{graphicx}
\definecolor{high}{rgb}{0.9312692223325372, 0.8201921796082118, 0.7971480974663592}
\definecolor{medium}{rgb}{0.7840440880599453, 0.5292660544265891, 0.6200568926941761}
\definecolor{low}{rgb}{0.5151069036855755, 0.29801047535056074, 0.49050619139300705}
\begin{document}
\title{Primary Tumor and Inter-Organ Augmentations for Supervised Lymph Node Colon Adenocarcinoma Metastasis Detection} 
\titlerunning{Augmentations for Supervised Colon Cancer Metastasis Detection}
\author{Apostolia Tsirikoglou\inst{1}\orcidID{0000-0003-0298-937X}\and \\Karin Stacke\inst{1,3}\orcidID{0000-0003-1066-3070}\and Gabriel	Eilertsen\inst{1,2}\orcidID{0000-0002-9217-9997}\and Jonas Unger\inst{1,2}\orcidID{0000-0002-7765-1747}}
\authorrunning{A. Tsirikoglou et al.}
\institute{Department of Science and Technology, Linkoping University, Sweden \and
Center for Medical Image Science and Visualization, Linkoping University, Sweden \and
Sectra AB, Sweden}
\maketitle              
\begin{abstract}
The scarcity of labeled data is a major bottleneck for developing accurate and robust deep learning-based models for histopathology applications. The problem is notably prominent for the task of metastasis detection in lymph nodes, due to the tissue's low tumor-to-non-tumor ratio, resulting in labor- and time-intensive annotation processes for the pathologists. This work explores alternatives on how to augment the training data for colon carcinoma metastasis detection when there is limited or no representation of the target domain. Through an exhaustive study of cross-validated experiments with limited training data availability, we evaluate both an inter-organ approach utilizing already available data for other tissues, and an intra-organ approach, utilizing the primary tumor. Both these approaches result in little to no extra annotation effort. Our results show that these data augmentation strategies can be an efficient way of increasing accuracy on metastasis detection, but fore-most increase robustness.

\end{abstract}
\input{introduction}
\input{background}
\input{data_experiments}
\input{results}
\input{conclusion}

%
%
%
\bibliographystyle{splncs04}
\bibliography{references}
\end{document}


\title{Primary Tumor and Inter-Organ Augmentations for Supervised Lymph Node Colon Adenocarcinoma Metastasis Detection \\ \vspace{.8cm}\large Supplementary material
} 
\titlerunning{Augmentations for Supervised Colon Cancer Metastasis Detection}
\author{Apostolia Tsirikoglou\inst{1}\orcidID{0000-0003-0298-937X}\and \\Karin Stacke\inst{1,3}\orcidID{0000-0003-1066-3070}\and Gabriel	Eilertsen\inst{1,2}\orcidID{0000-0002-9217-9997}\and Jonas Unger\inst{1,2}\orcidID{0000-0002-7765-1747}}
\authorrunning{A. Tsirikoglou et al.}
\institute{Department of Science and Technology, Linkoping University, Sweden \and
Center for Medical Image Science and Visualization, Linkoping University, Sweden \and
Sectra AB, Sweden}
\maketitle              
\vspace{1cm}
This document supports the main paper by providing the full set of results from the experiments and information around their implementation. 
\section{Colon, breast and skin datasets} 

The colon adenocarcinoma dataset, used for the conducted experiments consists of $394$ hematoxylin and eosin (H\&E) whole slide images (WSIs), from which $155$ contain tumor annotations. The data come from two medical centers in Sweden (Link\"oping and G\"avle) and correspond to $37$ anonymized individual cases. The dataset contains primary tumor samples as well as lymph node tumor and non-tumor tissue. The WSIs were sampled using a random uniform grid with $128$ microns between the sample points. This corresponds to $256$ pixels when sampling at a resolution of $0.5$ microns (i.e., approximately $200$ times magnification). We set the patch size to 256$\times$256, meaning that the patches were sampled side-by-side without overlapping. In total, $269,054$ patches from non-tumor, primary tumor, and lymph node tumor tissue were extracted. Each patch was assigned the label based on the annotation of the center pixel in the patch. Number of patches per tissue label and medical center can be found in Table~\ref{tab:four_groups_eight_patients_new_split_revision}. 

The breast dataset consists of $50$ H\&E WSIs of sentinel lymph node tissue, out of which $49$ contain tumor annotations. These are coming from five different medical centers in The Netherlands, where each center contributes $10$ WSIs. The dataset corresponds to $43$ individual cases, and it was sampled following the same strategy as in the colon dataset resulting in $200,770$ extracted patches.

The skin dataset used for the described experiments consists of $96$ H\&E WSIs, where $34$ of them contain tumor annotations identified as basal cell carcinoma, squamous cell carcinoma, and squamous cell carcinoma in situ. These data correspond to $71$ individual cases. The non-tumor patches were extracted in the same way as the colon and breast data, while for the tumor patches the sampling was performed using a random uniform grid with $96$ microns between the sample points. This corresponds to $192$ pixels when sampling at a resolution of $0.5$ microns. The patch size is also set to $256\times256$, which means that the patches were sampled side-by-side with a $25\%$ overlap (64 pixels). In total, they were extracted $277,193$ tumor and non-tumor patches.

We performed a train/val/test split for all datasets. The train/test split was conducted on a number-of-patients basis keeping a similar ratio of $32/5$, $37/6$, and $61/10$ patients for colon, breast, and skin respectively. The train/val split was done over the training sets on a $90/10\%$ ratio. For reference, we provide in Table~\ref{tab:reference_datasets} the datasets sizes.

\begin{table}[th]
    \centering
    \begin{tabular}{rcccc}
         & \multicolumn{2}{c}{TRAIN} & \multicolumn{2}{c}{TEST} \\ \cmidrule(lr){2-3} \cmidrule(lr){4-5}
         & Tumor & Non-tumor & Tumor & Non-tumor \\
         \toprule
        Colon &  101,909 & 132,612 & 17,565 & 16,968 \\
        Breast & 62,805 & 110,011 & 12,935 & 15,019 \\
        Skin & 28,285 & 211,582 & 5,622 & 31,704
    \end{tabular}
    \caption{Number of patch images for the training and test sets of colon, breast and skin datasets.}
    \label{tab:reference_datasets}
\end{table}

\section{Sub-division of the colon dataset in groups}

To explore scenarios with limited training data and to cross-validate the results of our experiments, we created four sub-sets out of the colon cancer dataset, based on a patient-level split. Each group consists of eight patients, that each one of them contributes to one or more tissue types and classes (primary tumor tissue, lymph node tumor/non-tumor). The split size was decided to give an extreme minimum of few thousand images per group, compared to the $\sim100,000$ images per class of the colon dataset.

The primary tumor tissue is represented less frequently in the colon dataset. Therefore, and for the low-cost annotation scenario to be satisfied for all groups with an adequate amount of data, the main criterion for the patients split was for each group to have approximately the same amount of primary tumor patch images. We also made sure that each group and baseline experiment included the same number of patients from the two different medical centers (G\"avle and Link\"oping), as well as that patients with a high number of images did not over-dominate the experiment.

\begin{table}[t!]
    \centering
    \caption{Number of training patches of colon dataset sub-sets used to simulate low training data availability and cross-validate the experiments results. HIGH, MEDIUM, and LOW refer to the baseline experiments defined by the training set's annotation cost. The HIGH cost includes only lymph node tumor training data, the MEDIUM mostly primary tumor and a few lymph node tumor samples, and the LOW cost only primary tumor tissue. For all three baseline experiments, the non-tumor class consists of lymph node tissue. TRAIN and TEST refer to the colon full set's training and testing sets respectively.}
    \label{tab:four_groups_eight_patients_new_split_revision}
    \begin{adjustbox}{max width=.95\textwidth}
        \begin{tabular}{lrcccccccc}
         &  & \multicolumn{2}{c}{\multirow{2}{*}{tumor / non-tumor}} & \multicolumn{6}{c}{Link\"oping / G\"avle} \\ \cmidrule(lr){5-10}
         & \multirow{3}{*}{\rot{GROUP}} & & & \multicolumn{4}{c}{tumor} & \multicolumn{2}{c}{\multirow{2}{*}{non-tumor}} \\ \cmidrule(lr){3-4} \cmidrule(lr){5-8}
         & & & & \multicolumn{2}{c}{lymph} & \multicolumn{2}{c}{primary} & & \\ \cmidrule(lr){5-6} \cmidrule(lr){7-8} \cmidrule(lr){9-10}
         & & pre-bal. & post-bal. & pre-bal. & post-bal. & pre-bal. & post-bal. & pre-bal. & post-bal. \\
        \toprule
        \multirow{4}{*}{\rot{{\footnotesize \textcolor{high}{\textbf{HIGH}}}}} 
         & 0 & 19,077 / 24,772 & 19,077 / 19,080 & \multicolumn{2}{c}{20 / 19,057} & \multicolumn{2}{c}{--} & 20,466 / 4,306 & 14,774 / 4,306 \\
         & 1 & 10,585 / 45,303 & 10,585 / 10,585 & \multicolumn{2}{c}{2,851 / 7,734} & \multicolumn{2}{c}{--} & 34,991 / 10,312 & 4,234 / 6,351 \\
         & 2 & 22,941 / 28,005 & 22,941 / 22,913 & \multicolumn{2}{c}{1,722 / 21,219} & \multicolumn{2}{c}{--} & 17,797 / 10,208 & 12,916 / 9,997 \\
         & 3 & 26,342 / 34,532 & 26,342 / 26,346 & \multicolumn{2}{c}{10,171 / 16,171} & \multicolumn{2}{c}{--} & 21,063 / 13,469 & 16,632 / 9,714 \\ \cmidrule(lr){2-10}
        \multirow{4}{*}{\rot{{\footnotesize \textcolor{medium}{\textbf{MEDIUM}}}}} 
         & 0 & 16,313 / 10,346 & 7,184 / 7,184 & 20 / 11,504 & 20 / 2,375 & \multicolumn{2}{c}{0 / 4,789} & 20,466 / 4,306 & 3,045 / 4,139 \\
         & 1 & 11,173 / 38,230 & 8,771 / 8,772 & 2,851 / 2,475 & 1,462 / 1,462 & \multicolumn{2}{c}{878 / 4,969} & 34,991 / 10,312 & 4,386 / 4,386 \\
         & 2 & 9,525 / 19,778 & 8,512 / 8,503 & 1,559 / 2,292 & 1,419 / 1,419 & \multicolumn{2}{c}{537 / 5,137} & 17,797 / 10,208 & 3,723 / 4,789 \\
         & 3 & 26,511 / 25,735 & 9,982 / 9,982 & 10,129 / 9,728 & 1,664 / 1,664 & \multicolumn{2}{c}{4,578 / 2,076} & 21,063 / 13,469 & 4,991 / 4,991 \\ \cmidrule(lr){2-10}
        \multirow{4}{*}{\rot{{\footnotesize \textcolor{low}{\textbf{LOW}}}}} 
         & 0 & 4,789 / 24,772  & 4,789 / 4,789 & \multicolumn{2}{c}{--} & \multicolumn{2}{c}{0 / 4,789} & 20,466 / 4,306 & 650 / 4,139 \\
         & 1 & 5,847 / 45,303  & 5,847 / 5,848 & \multicolumn{2}{c}{--} & \multicolumn{2}{c}{878 / 4,969} & 34,991 / 10,312 & 2,924 / 2,924 \\
         & 2 & 5,674 / 28,005 & 5,674 / 5,668 & \multicolumn{2}{c}{--} & \multicolumn{2}{c}{537 / 5,137} & 17,797 / 10,208 & 2,837 / 2,837 \\
         & 3 & 6,654 / 34,532 & 6,654 / 6,654  & \multicolumn{2}{c}{--} & \multicolumn{2}{c}{4,578 / 2,076} & 21,063 / 13,469 & 3,327 / 3,327 \\ 
        \midrule
        \rot{{\footnotesize TRAIN}} &  & 101,909 / 132,612 & 101,909 / 101,909 & \multicolumn{2}{c}{14,764 / 64,181} & \multicolumn{2}{c}{5,993 / 16,971} & 94,317 / 38,295 & 63,614/38,295 \\ \cmidrule(lr){1-10} \\ 
        \rot{{\footnotesize TEST}} &  & 17,565 / 16,968 & 13,167 / 16,968 & \multicolumn{2}{c}{980 / 12,187} & 0 / 4,398 & 0 / 0 & \multicolumn{2}{c}{6,410 / 10,558} \\
    \end{tabular}
    \end{adjustbox}
\end{table}

In the high-cost annotation case, each group utilizes all the available lymph node tumor tissue (coming from at least six patients per group), along with the same amount of non-tumor lymph node tissue. The latter do not necessarily come from the same patients that provide the lymph node tumor samples. Medium cost case leverages all the per group available primary tumor patches (coming from at least four patients) along with lymph node tumor tissue equal to half of the size of primary tumor samples (coming from only two patients). In this case, the two patients that provide the lymph tumor samples, also supply the non-tumor lymph node class. Finally, for the low-cost scenario, only the primary tumor is used (with no representation of the target tumor domain), while the non-tumor lymph node class is formed by two patients per group. For all three annotation cost experiments, the non-tumor balancing to the size of the tumor class was conducted as random patches selection from either all or the specified patients. 

Table~\ref{tab:four_groups_eight_patients_new_split_revision} presents the colon dataset split per group and annotation cost baseline experiment. The number of patches is given in the total per case tumor/non-tumor ratio, as well as in a medical site and per tissue type detailed view pre- and post- class balancing. 

\section{Results}
\begin{figure}[ht]
\centering
    \includegraphics[angle=90,width=\textwidth]{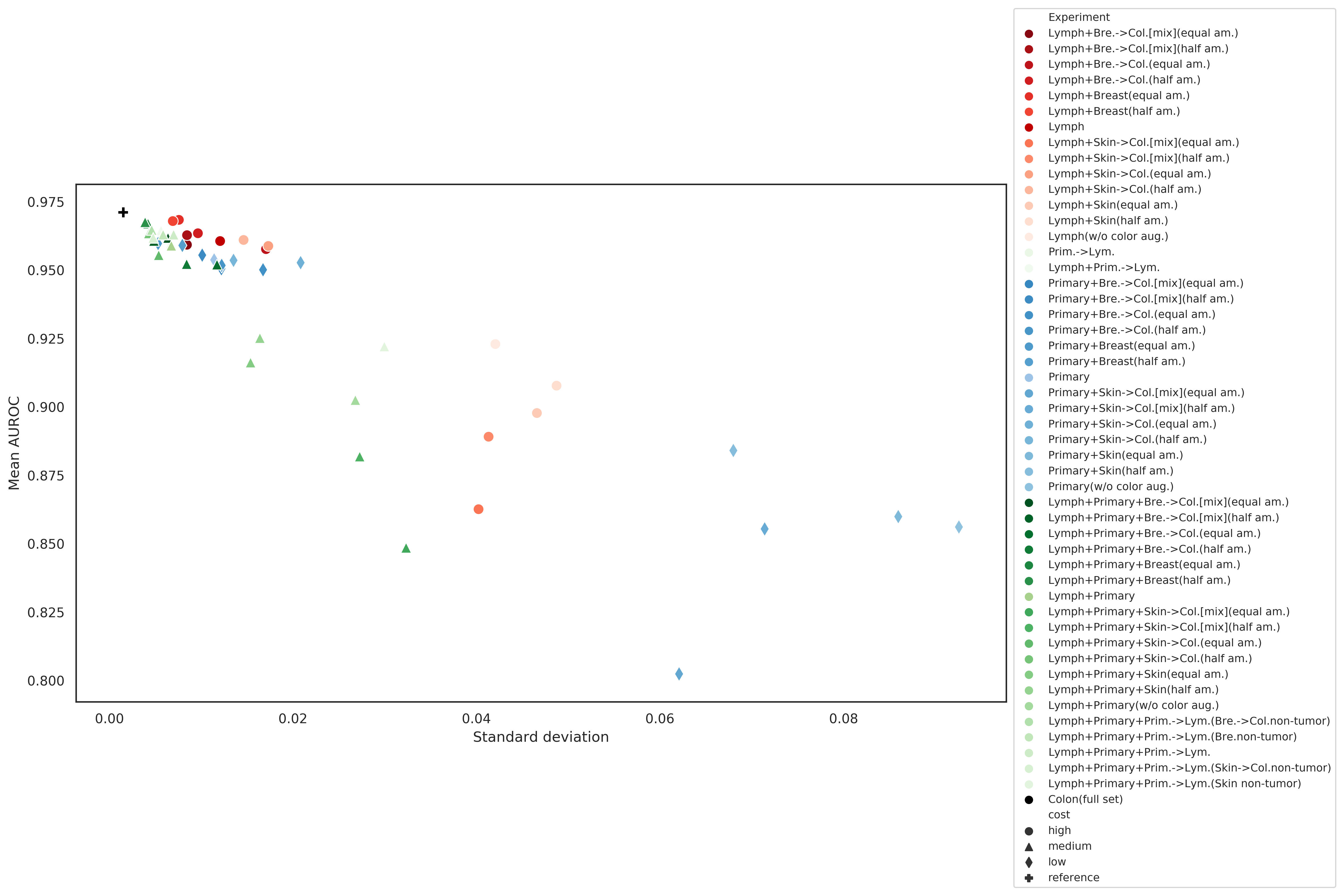}
    \caption{Mean AUROC against the standard deviation computed over the four  sub-sets' performance for five trainings per sub-set, for all tested experiments.}
    \label{fig:auroc_stddev_legend}
\end{figure}

\begin{figure}[ht]
\centering
    \includegraphics[angle=90,height=\textheight]{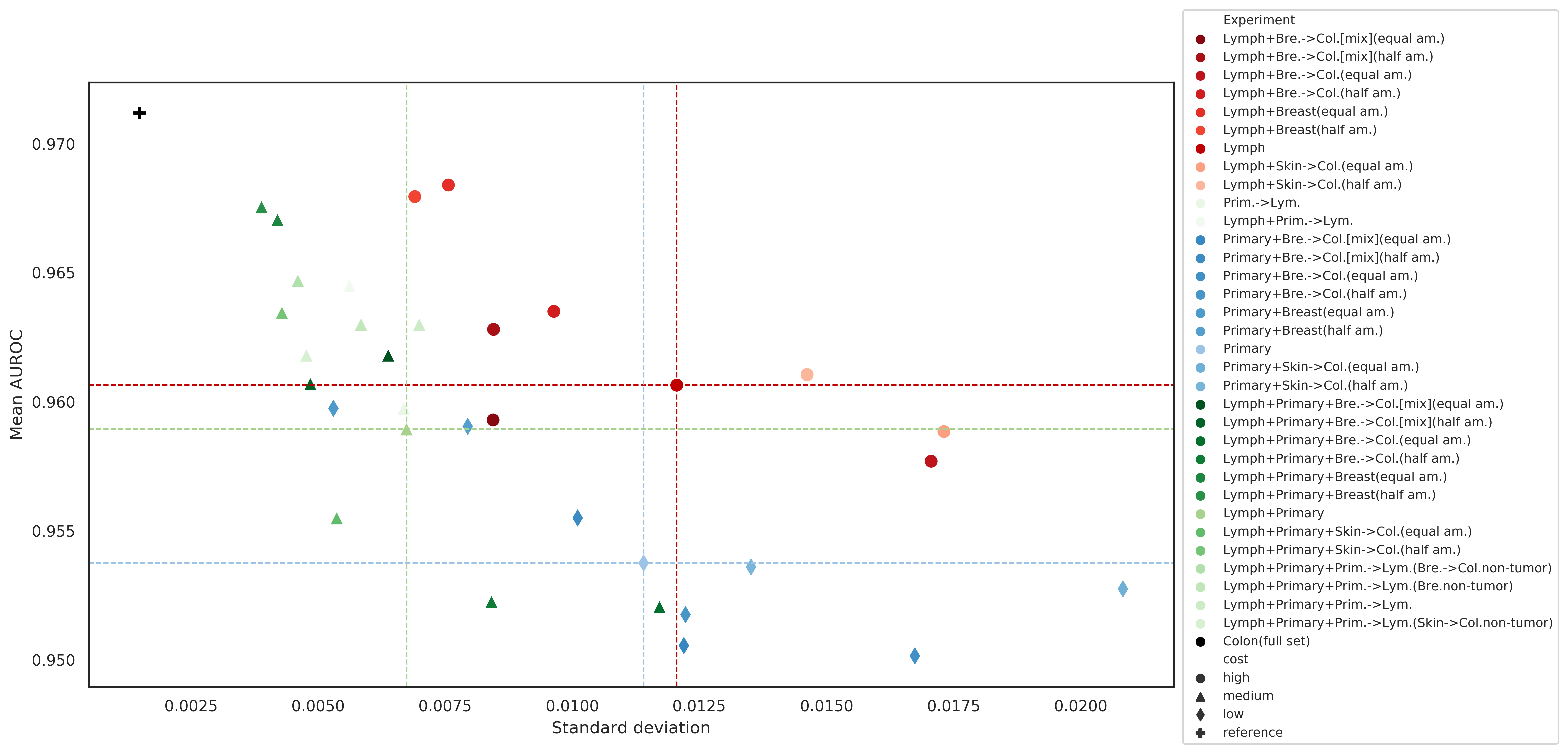}
    \caption{Mean AUROC against the standard deviation computed over the four  sub-sets' performance for five trainings per sub-set, for the best performing scenarios and the corresponding baselines.}
    \label{fig:best_auroc_stddev_legend}
\end{figure}

\begin{figure}
    \centering
    \includegraphics[width=\textwidth]{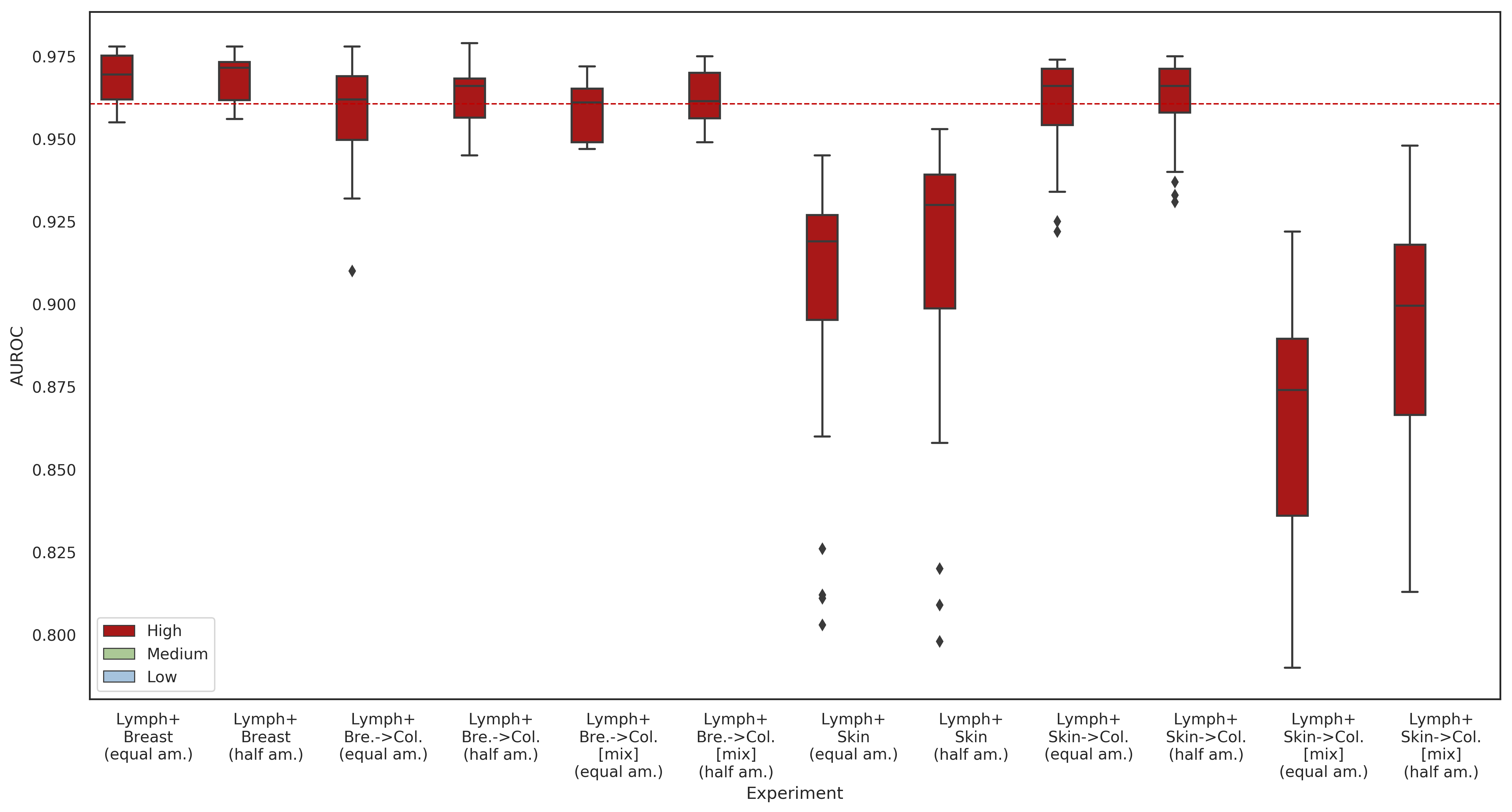} \\
    \includegraphics[width=\textwidth]{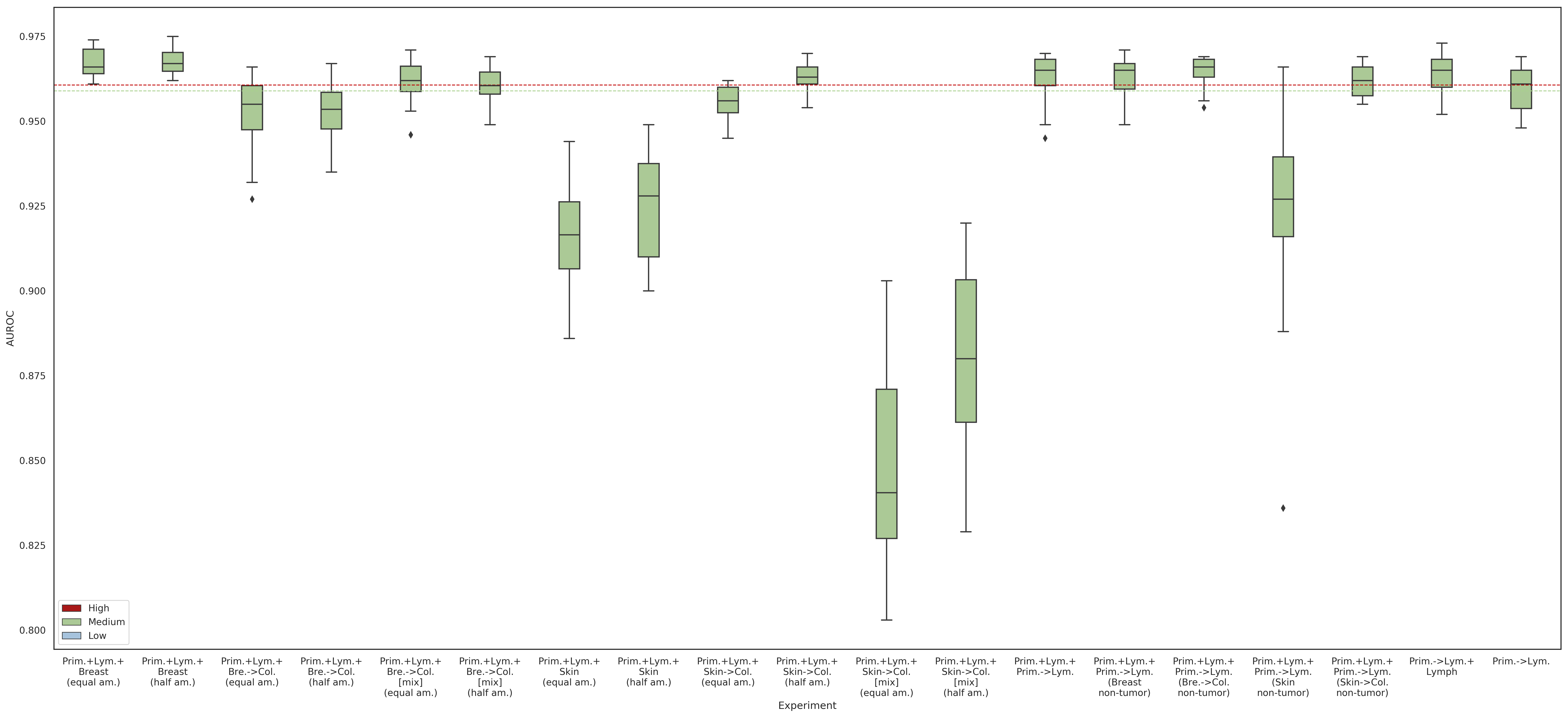} \\
    \includegraphics[width=\textwidth]{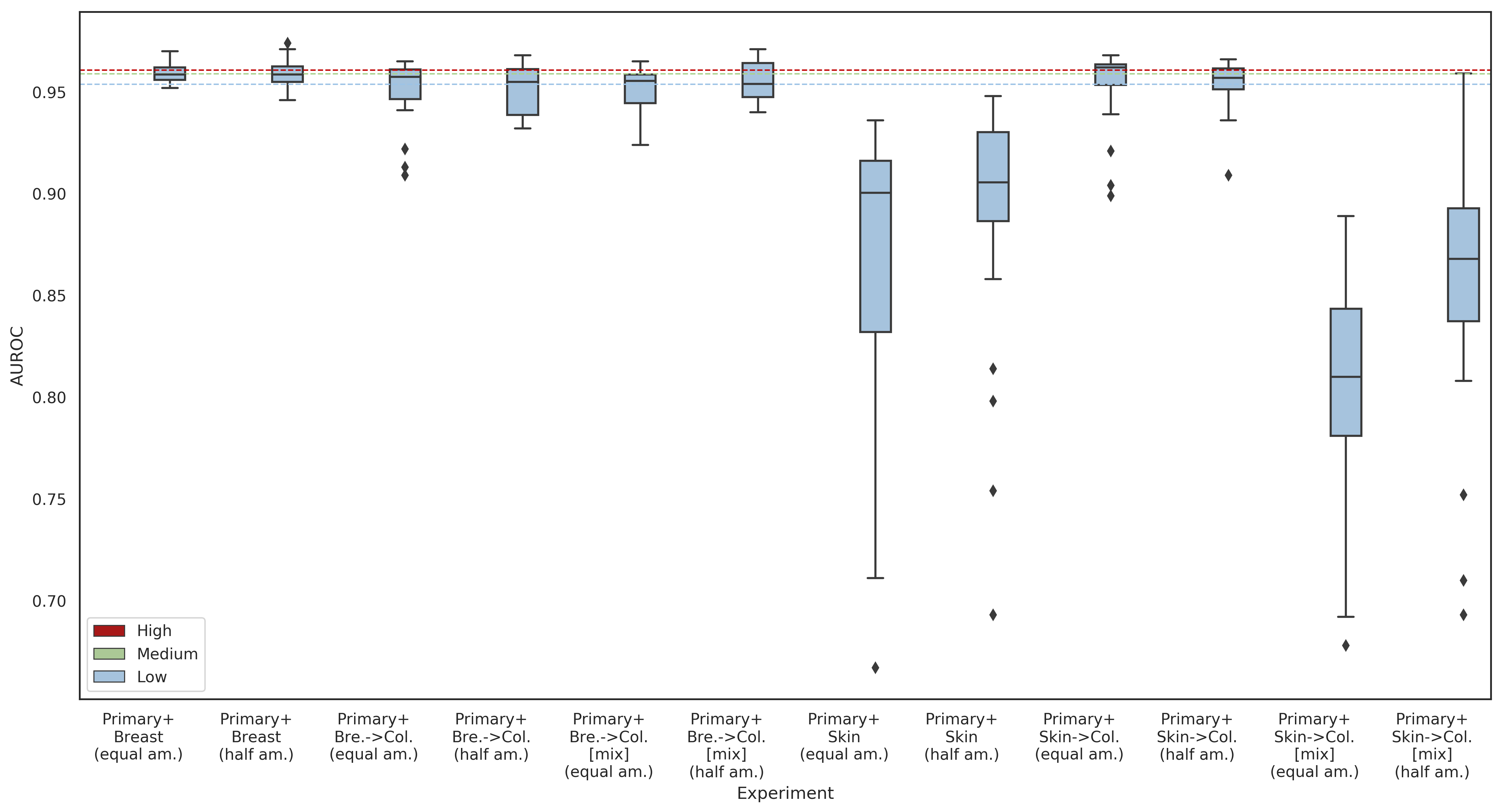}
    \caption{AUROC for high (top), medium (middle) and low (bottom) cost scenarios against the mean perfomance for the baseline annotation effort scenarios(dashed lines), for all the tested augmentation combinations and strategies. The boxes show the quartiles of the perfomance results per experiment, while the whiskers extend to show the rest of the distribution, except for the outliers (diamond markers).}
    \label{fig:all_cost_exp}
\end{figure}

Here we complement the most central results presented in the  main paper with an account for the full set of experiments conducted in the study and additional plots describing the performance in the different scenarios and for different augmentations. Table~\ref{tab:experiments_mean_acc_auroc_stddev} shows the \textbf{AUROC} (Area Under the ROC curve) for all experiments. This is also shown in the plot in Figure~\ref{fig:auroc_stddev_legend} where the mean AUROC is plotted against the standard deviation computed over the four sub-sets' performance for five trainings per sub-set. The zoom in in Figure~\ref{fig:best_auroc_stddev_legend} shows a selection of the best performing scenarios and the corresponding baselines. From top to bottom, the box plots in Figure~\ref{fig:all_cost_exp} show the AUROC for high, medium and low cost scenarios against the mean performance for the baseline annotation effort scenarios, i.e., with no augmentations (dashed line), for the different data/augmentation combinations.
The boxes show the quartiles of the perfomance results per experiment, while the whiskers extend to show the rest of the distribution, except for the outliers (diamond markers).

In the experiments presented in Table~\ref{tab:experiments_mean_acc_auroc_stddev} and Figures~\ref{fig:auroc_stddev_legend},~\ref{fig:best_auroc_stddev_legend} and~\ref{fig:all_cost_exp} left to right arrow ($\rightarrow$) denote data domain adaptation. The suffix [mix] stands for Cycle-GAN transformations in a class-agnostic fashion. For example, Bre.$\rightarrow$Col.[mix] means that the breast tissue data were transformed to the target domain without performing per-class adaptations, while Bre.$\rightarrow$Col. means that tumor and non-tumor breast tissue data were transformed by separate Cycle-GANs to tumor and non-tumor colon tissue data respectively. Moreover, we include to the notation the augmented set's size in relation to the baseline training set number of patches; (equal am.) stands for equal amount of added images to the baseline train set, while (half am.) for half the amount.

\begin{figure}
    \centering
    \includegraphics[width=\textwidth,height=\textheight, keepaspectratio]{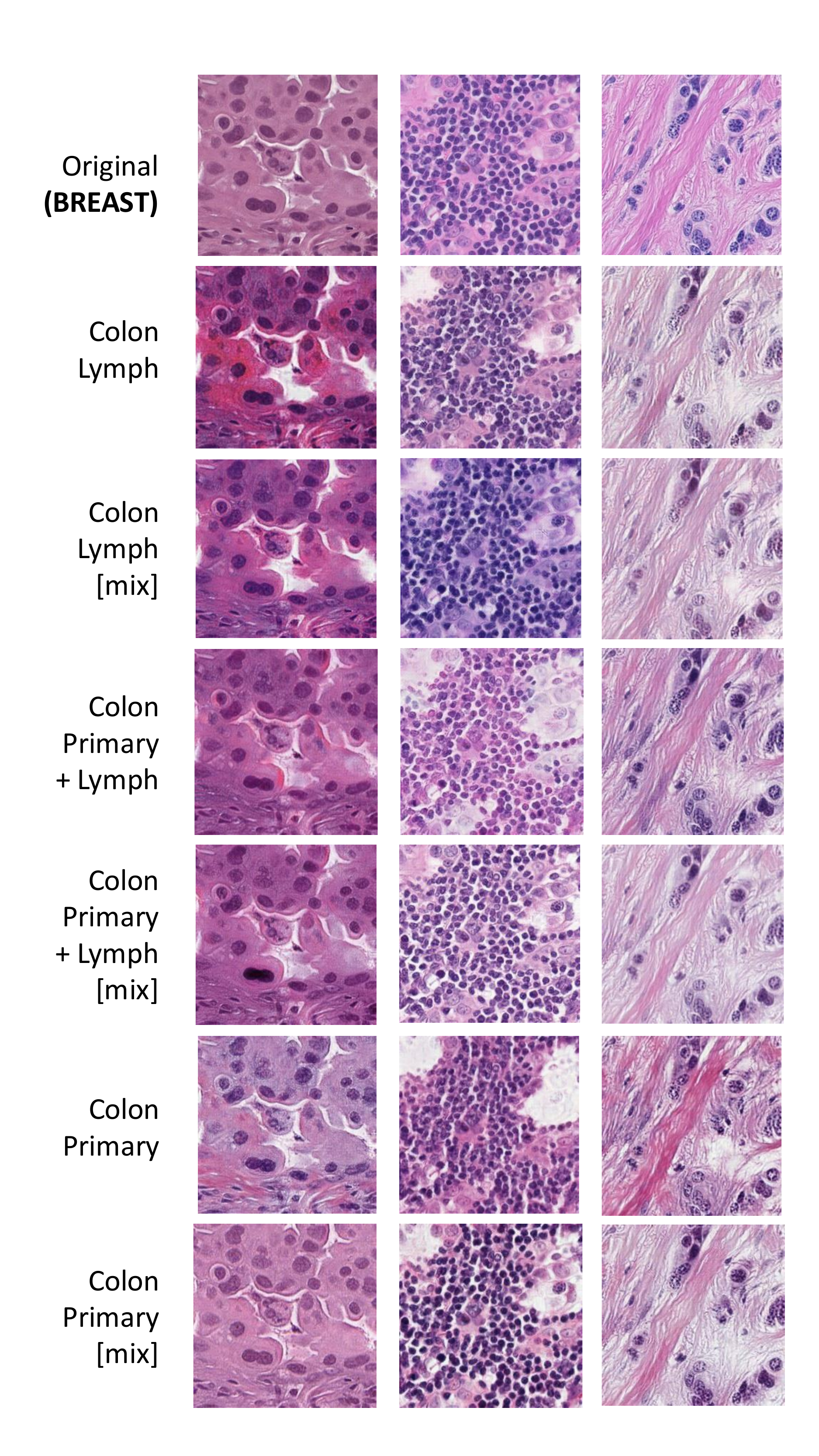}
    \caption{Data domain adaptation to the colon target domain for three example tumor patches of \textbf{breast} tissue utilizing image-to-image translation. The transformation differs depending on the colon data feeding the Cycle-GAN, as well as if whether two Cycle-GANs were trained separately for each class, or one Cycle-GAN was trained jointly for tumor and non-tumor tissue data (suffix [mix]).}
    \label{fig:data_domain_adaptation_breast}
\end{figure}

\begin{figure}
    \centering
    \includegraphics[width=\textwidth,height=\textheight, keepaspectratio]{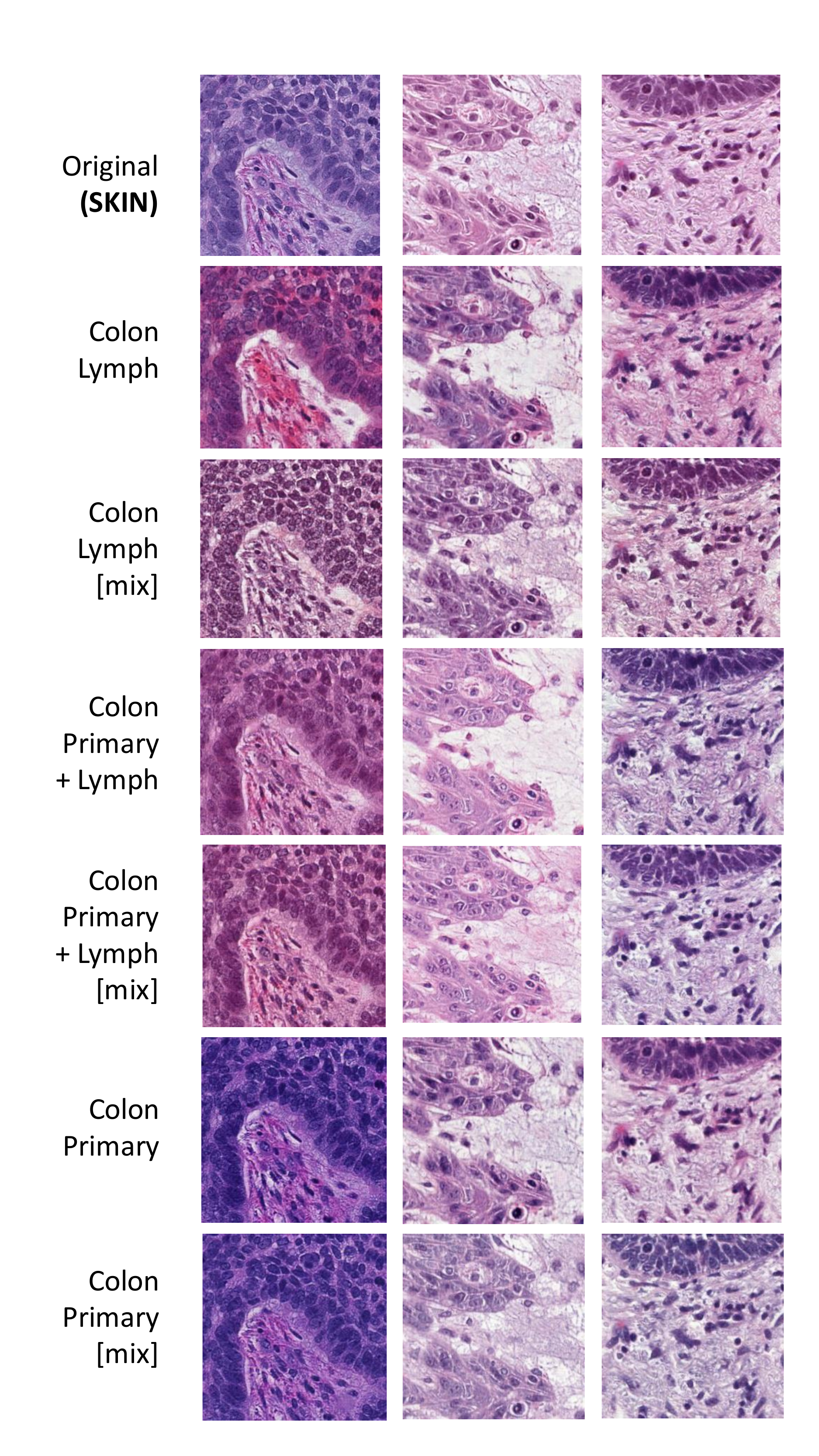}
    \caption{Data domain adaptation to the colon target domain for three example tumor patches of \textbf{skin} tissue utilizing image-to-image translation. The transformation differs depending on the colon data feeding the Cycle-GAN, as well as if whether two Cycle-GANs were trained separately for each class, or one Cycle-GAN was trained jointly for tumor and non-tumor tissue data (suffix [mix]).}
    \label{fig:data_domain_adaptation_skin}
\end{figure}

Finally, Figures~\ref{fig:data_domain_adaptation_breast} and~\ref{fig:data_domain_adaptation_skin} provide examples of data domain adaptation for breast and skin tumor tissue respectively for visual inspection. The image-to-image translation differs for the various colon data available for the Cycle-GANs training, as well as the training approach. We test both training separate Cycle-GANs for each of the tumor and non-tumor classes, and train one joint network for both classes.

\newpage

\begin{table}[h!]
    \centering
    \caption{Mean \textbf{patch accuracy} and \textbf{AUROC} for the experiments along with the standard deviation between the sub-sets' and identical trainings performances. All classifiers have trained with color augmentation, if not mentioned otherwise.}
    \label{tab:experiments_mean_acc_auroc_stddev}
    \begin{adjustbox}{max width=\textwidth}
        \begin{tabular}{rcccc}
         & \multicolumn{3}{c}{Mean Accuracy$\pm$stddev} & \\
        \cmidrule(lr){2-4}
        Experiment & lymph tumor & non-tumor & total & Mean AUROC$\pm$stddev \\
        \toprule
        Lymph(w/o color aug.) & 0.8733$\pm$0.0974 & 0.9727$\pm$0.0183 & 0.9293$\pm$0.0327 & 0.9230$\pm$0.0421\\
        Lymph + Primary(w/o color aug.) & 0.8588$\pm$0.0787 & 0.9463$\pm$0.04220 & 0.9081$\pm$0.0144 & 0.9026$\pm$0.0268 \\
        Primary(w/o color aug.) & 0.7464$\pm$0.1834 & 0.9657$\pm$0.0272 & 0.8699$\pm$0.0853 & 0.8562$\pm$0.0926 \\
        \cmidrule(lr){1-5}
        Lymph & 0.9541$\pm$0.0231 & 0.9671$\pm$0.0176 & 0.9614$\pm$0.0114 & 0.9607$\pm$0.0121 \\
        Lymph + Primary & 0.9673$\pm$0.0126 & 0.9507$\pm$0.0253 & 0.9580$\pm$0.0091 & 0.9590$\pm$0.0067 \\
        Primary & 0.9501$\pm$0.0139 & 0.9573$\pm$0.0228 & 0.9542$\pm$0.0114 & 0.9538$\pm$0.0114 \\
        \cmidrule(lr){1-5}
        Lymph + Breast (equal am.) & 0.9601$\pm$0.0148 & 0.9768$\pm$0.0097 & 0.9695$\pm$0.0074 & 0.9684$\pm$0.0076 \\
        Lymph + Breast (half am.) & 0.9558$\pm$0.0149 & 0.9801$\pm$0.0083 & 0.9695$\pm$0.0065 & 0.9680$\pm$0.0069 \\
        Lymph + Bre.$\rightarrow$Col. (equal am.) & 0.9402$\pm$0.0244 & 0.9752$\pm$0.0175 & 0.9599$\pm$0.0154 & 0.9577$\pm$0.0171 \\
        Lymph + Bre.$\rightarrow$Col. (half am.) & 0.9509$\pm$0.0136 & 0.9761$\pm$0.0141 & 0.9651$\pm$0.0097 & 0.9635$\pm$0.0096 \\
        Lymph + Bre.$\rightarrow$Col.[mix] (equal am.) & 0.9382$\pm$0.0106 & 0.9803$\pm$0.0077 & 0.9619$\pm$0.0068 & 0.9593$\pm$0.0084 \\
        Lymph + Bre.$\rightarrow$Col.[mix] (half am.) & 0.9455$\pm$0.0138 & 0.9802$\pm$0.0096 & 0.9650$\pm$0.0073 & 0.9628$\pm$0.0085 \\
        Lymph + Skin (equal am.) & 0.8235$\pm$0.0947 & 0.9719$\pm$0.0082 & 0.9071$\pm$0.0447 & 0.8978$\pm$0.0466 \\
        Lymph + Skin (half am.) & 0.8364$\pm$0.1014 & 0.9791$\pm$0.0051 & 0.9167$\pm$0.04455 & 0.9078$\pm$0.0487 \\
        Lymph + Skin$\rightarrow$Col. (equal am.) & 0.9473$\pm$0.0332 & 0.9706$\pm$0.0090 & 0.9604$\pm$0.0174 & 0.9589$\pm$0.0173 \\
        Lymph + Skin$\rightarrow$Col. (half am.) & 0.9481$\pm$0.0286 & 0.9739$\pm$0.0095 & 0.9626$\pm$0.0145 & 0.9611$\pm$0.0146 \\
        Lymph + Skin$\rightarrow$Col.[mix] (equal am.) & 0.7573$\pm$0.0822 & 0.9681$\pm$0.0150 & 0.8760$\pm$0.0354 & 0.8627$\pm$0.0402 \\
        Lymph + Skin$\rightarrow$Col.[mix] (half am.) & 0.8050$\pm$0.0861 & 0.9732$\pm$0.0117 & 0.8997$\pm$0.0396 & 0.8892$\pm$0.0413 \\
        \cmidrule(lr){1-5}
        Lymph +  Primary + Breast (equal am.) & 0.9665$\pm$0.0103 & 0.9676$\pm$0.0101 & 0.9671$\pm$0.0042 & 0.9671$\pm$0.0042 \\
        Lymph + Primary + Breast (half am.) & 0.9669$\pm$0.0128 & 0.9681$\pm$0.0145 & 0.9676$\pm$0.0044 & 0.9676$\pm$0.0039 \\
        Lymph + Primary + Bre.$\rightarrow$Col. (equal am.) & 0.9675$\pm$0.0158 & 0.9367$\pm$0.0327 & 0.9501$\pm$0.0123 & 0.9521$\pm$0.0117 \\
        Lymph + Primary + Bre.$\rightarrow$Col. (half am.) & 0.9667$\pm$0.0213 & 0.9377$\pm$0.0297 & 0.9503$\pm$0.0096 & 0.9523$\pm$0.0084 \\
        Lymph + Primary + Bre.$\rightarrow$Col.[mix] (equal am.) & 0.9691$\pm$0.0125 & 0.9545$\pm$0.0159 & 0.9609$\pm$0.0048  &0.9618$\pm$0.0064 \\
        Lymph + Primary + Bre.$\rightarrow$Col.[mix] (half am.) & 0.9511$\pm$0.0155 & 0.9704$\pm$0.0125 & 0.9620$\pm$0.0032 & 0.9607$\pm$0.0048 \\
        Lymph + Primary + Skin (equal am.) & 0.8649$\pm$0.0083 & 0.9677$\pm$0.0099 & 0.9228$\pm$0.0075  & 0.9163$\pm$0.0154 \\
        Lymph + Primary + Skin (half am.) & 0.8762$\pm$0.0193 & 0.9743$\pm$0.0083 & 0.9314$\pm$0.0087 & 0.9252$\pm$0.0164 \\
        Lymph + Primary + Skin$\rightarrow$Col. (equal am.) & 0.9543$\pm$0.0194 & 0.9567$\pm$0.0213 & 0.9557$\pm$0.0058 & 0.9555$\pm$0.0054 \\
        Lymph + Primary + Skin$\rightarrow$Col. (half am.) & 0.9610$\pm$0.0162 & 0.9661$\pm$0.0132 & 0.9638$\pm$0.0038 & 0.9635$\pm$0.0043 \\
        Lymph + Primary + Skin$\rightarrow$Col.[mix] (equal am.) & 0.7290$\pm$0.0503 & 0.9682$\pm$0.0084 & 0.8637$\pm$0.0264 & 0.8486$\pm$0.0323 \\
        Lymph + Primary + Skin$\rightarrow$Col.[mix] (half am.) & 0.7934$\pm$0.0113 & 0.9705$\pm$0.0157 & 0.8931$\pm$0.0104 & 0.8819$\pm$0.0273 \\
        \cmidrule(lr){1-5} 
        Prim.$\rightarrow$Lym. & 0.9486$\pm$0.0222 & 0.9709$\pm$0.0166 & 0.9611$\pm$0.0058 & 0.9598$\pm$0.0067 \\
        Lymph + Prim.$\rightarrow$Lym. & 0.9556$\pm$0.0144 & 0.9733$\pm$0.0117 & 0.9656$\pm$0.0046 & 0.9645$\pm$0.0056 \\
        \cmidrule(lr){1-5}
        Lymph + Primary + Prim.$\rightarrow$Lym. & 0.9535$\pm$0.0154 & 0.9725$\pm$0.0119 & 0.9642$\pm$0.0064 & 0.9630$\pm$0.0070 \\
        Lymph + Primary + Prim.$\rightarrow$Lym. + Breast non-tumor & 0.960$\pm$0.0143 & 0.9656$\pm$0.0199 & 0.9631$\pm$0.0055 & 0.9630$\pm$0.0058\\
        Lymph + Primary + Prim.$\rightarrow$Lym. + Bre.$\rightarrow$Col. non-tumor & 0.9570$\pm$0.0090 & 0.9723$\pm$0.0120 & 0.9656$\pm$0.0052 &0.9647$\pm$0.0046 \\
        Lymph + Primary + Prim.$\rightarrow$Lym. + Skin non-tumor & 0.8719$\pm$0.0354 & 0.9741$\pm$0.0165 & 0.9295$\pm$0.0168 & 0.9221$\pm$0.030 \\
        Lymph + Primary + Prim.$\rightarrow$Lym. + Skin$\rightarrow$Col. non-tumor & 0.9576$\pm$0.0185 & 0.9668$\pm$0.0177 & 0.9628$\pm$0.0031 &0.9618$\pm$0.0048 \\
        \cmidrule(lr){1-5}
        Primary + Breast (equal am.) & 0.9514$\pm$0.0150 & 0.9681$\pm$0.0146 & 0.9608$\pm$0.0053 & 0.9598$\pm$0.0053\\
        Primary + Breast (half am.) & 0.9498$\pm$0.0176 & 0.9683$\pm$0.0133 & 0.9603$\pm$0.0068 & 0.9591$\pm$0.0079\\
        Primary + Bre.$\rightarrow$Col. (equal am.) & 0.9367$\pm$0.0193 & 0.9637$\pm$0.0166 & 0.9519$\pm$0.0157 & 0.9502$\pm$0.0167 \\
        Primary + Bre.$\rightarrow$Col. (half am.) & 0.9432$\pm$0.0210 & 0.9603$\pm$0.0212 & 0.9528$\pm$0.0129 & 0.9518$\pm$0.0122 \\
        Primary + Bre.$\rightarrow$Col.[mix] (equal am.) & 0.9285$\pm$0.0189 & 0.9726$\pm$0.0130 & 0.9534$\pm$0.0112 & 0.9506$\pm$0.0122 \\
        Primary + Bre.$\rightarrow$Col.[mix] (half am.) & 0.9409$\pm$0.0261 & 0.9702$\pm$0.0141 & 0.9574$\pm$0.0087 & 0.9555$\pm$0.0101 \\
        Primary + Skin (equal am.) & 0.7534$\pm$0.1691 & 0.9665$\pm$0.0207 & 0.8734$\pm$0.0843 & 0.8600$\pm$0.0860 \\
        Primary + Skin (half am.) & 0.7967$\pm$0.1258 & 0.9715$\pm$0.0158 & 0.8951$\pm$0.0614 & 0.8841$\pm$0.0680 \\
        Primary + Skin$\rightarrow$Col. (equal am.) & 0.9341$\pm$0.0372 & 0.9714$\pm$0.0086 & 0.9551$\pm$0.0187 & 0.9528$\pm$0.0208 \\
        Primary + Skin$\rightarrow$Col. (half am.) & 0.9399$\pm$0.0201 & 0.9671$\pm$0.0140 & 0.9552$\pm$0.0117 & 0.9536$\pm$0.0135 \\
        Primary + Skin$\rightarrow$Col.[mix] (equal am.) & 0.6345$\pm$0.1147 & 0.9705$\pm$0.0107 & 0.8237$\pm$0.0545 & 0.8025$\pm$0.0621 \\
        Primary + Skin$\rightarrow$Col.[mix] (half am.) & 0.7386$\pm$0.0992 & 0.9722$\pm$0.0104 & 0.8701$\pm$0.0488 & 0.8555$\pm$0.0714 \\
        \midrule
        Colon full set(w/o color aug.) & 0.9608$\pm$0.0075 & 0.9814$\pm$0.0063 & 0.9724$\pm$0.0014 & 0.9712$\pm$0.0015
    \end{tabular}
    \end{adjustbox}
\end{table}

\newpage

%% file: introduction.tex
\section{Introduction}

Colon cancer is the third most common cancer type in the world, where the majority of the cases are classified as adenocarcinoma \cite{wild_2020}. Along with grading the primary tumor, assessment of the spread of the tumor to regional lymph nodes is an important prognostic factor \cite{compton_2000}. The pathologist is therefore required to not only assess the primary tumor but in high resolution, scan multiple lymph node sections, a task that is both challenging and time-consuming. Deep learning-based methods could be of use in assisting the pathologist, as they have shown great success for other histopathology applications \cite{Serag_2019}. However, a significant challenge is the need for large, annotated datasets, which in the case of lymph node metastasis detection is heightened due to the low tumor-to-non-tumor ratio in the tissue, and the annotation expertise needed. 
In this paper, we study how data with lower acquisition and annotation cost can be used to augment the training dataset, thus reducing the need for a large cohort size of the target lymph node metastasis data. We explore this using \textit{inter-organ augmentations}, i.e., utilizing data from different organs from existing public datasets. Leveraging the uniformity across staining, scanning, and annotation protocols, we investigate how potential similarities and differences across tissue and cancer types can be useful for the target application. In addition to the inter-organ augmentations, we also consider \emph{intra-organ augmentations}, by using data from the primary tumor. Gathering labeled data from the primary tumor requires little extra work, as tissue samples of it typically are acquired in conjunction with the lymph node sections, and the high tumor-to-non-tumor ratio allows for faster annotations. Furthermore, we investigate three different data availability scenarios based on annotation cost (in terms of time and effort). 

In summary, we present the following set of contributions:
\begin{itemize}
\item The first large-scale study on inter-organ data augmentations in digital pathology for metastasis detection. This includes a rigorous experimental setup of different combinations of inter- and intra-organ training data. We test both direct augmentation between organs, as well as transformed data by means of Cycle-GAN~\cite{Zhu_2017} in order to align the source images with the target domain.
\item We measure the impact on performance of lymph node colon tumor metastasis detection in three different scenarios, each representing a different effort/cost in terms of gathering and annotating the data.
\item In addition to inter-organ augmentation, we show how intra-organ augmentation, using data from the primary tumor, can increase robustness for detection on lymph node data.
\end{itemize}

The results point to how inter-organ data augmentations can be an important tool for boosting accuracy, but fore-mostly for increasing the robustness of deep pathology applications. How to make the best use of source organ data depends on its similarity to the target organ, where more similar data results in no or detrimental impact on performance together with Cycle-GAN transformed images, whereas the opposite is true for dissimilar data. Finally, we highlight the importance of making use of data from the primary tumor, as a low-effort strategy for increasing the robustness of lymph node metastasis detection.

%% file: background.tex
\section{Related work}

A number of deep learning-based methods have previously been presented for metastases detection, primarily facilitated by the CAMELYON16 and -17 challenges~\cite{ehteshamibejnordi_2017,bandi_2019}, where large datasets of whole-slide images of sentinel lymph node sections for breast cancer metastases were collected and made publicly available. As these types of large datasets are not available for all tissue and cancer types, different approaches have been taken to harness the data in other domains. These can be divided in to two, in many cases orthogonal categories: manipulation of the model, such as transfer learning \cite{Khan_2019,Xia_2018} and domain adaptation \cite{figueiraAdversarialBasedDomainAdaptation2020,Ren_2019}, 
and manipulation of the data, which is the focus of this paper.

Examples of augmentations that have shown successful for histopathology applications are geometric transformations (rotation, flipping, scaling), color jittering \cite{Tellez_2019,Stacke_2021}, and elastic deformations \cite{Karimi_2020}. Furthermore, methods using generative adversarial networks (GANs \cite{goodfellow_2014}) to synthetically generate data have proven useful \cite{Levine_2020,Krause_2021,Hou_2019,Brieu_2019}. 
In this work, we omit the step of generating synthetic data, and instead, investigate the possibility to augment the target dataset with 1) already existing publicly available datasets of other tissue types, and 2) same-distribution data with lower annotation cost.

Using the primary tumor for metastasis detection has been done in Zhou et al.~\cite{Zhou_2020} for preoperative investigation using ultrasound imaging, and in Lu et al. \cite{Lu_2020}, where metastatic tumor cells were used to find the primary source. To our knowledge, this is the first time the efficiency of using primary tumor data for metastatic cancer detection is investigated in histopathology. 

%% file: data_experiments.tex
\section{Method}\label{sec:method}
\begin{figure}[t!]
    \centering
    \includegraphics[width=.85\textwidth]{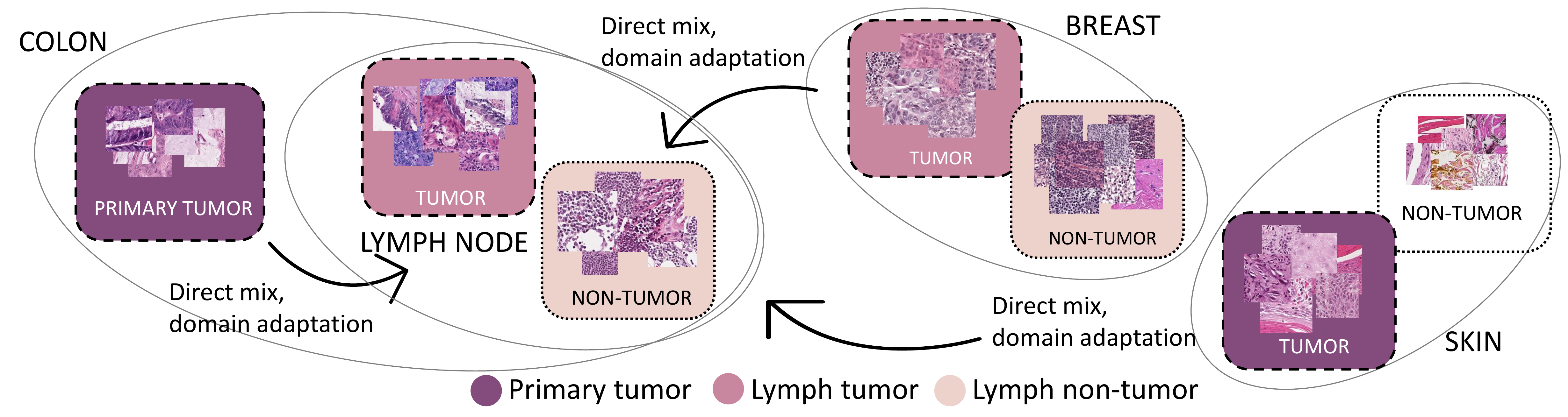}
    \caption{Same-distribution (colon primary tumor) as target, and inter-organ (breast and skin) augmentations. Both alternatives are explored either as data direct mix, or image synthesis as domain adaptations to the target distribution (lymph node colon adenocarcinoma metastasis).
    }
    \label{fig:data_exp_overview}
\end{figure}

\begin{figure}[t!]
    \centering
    \includegraphics[width=.9\textwidth]{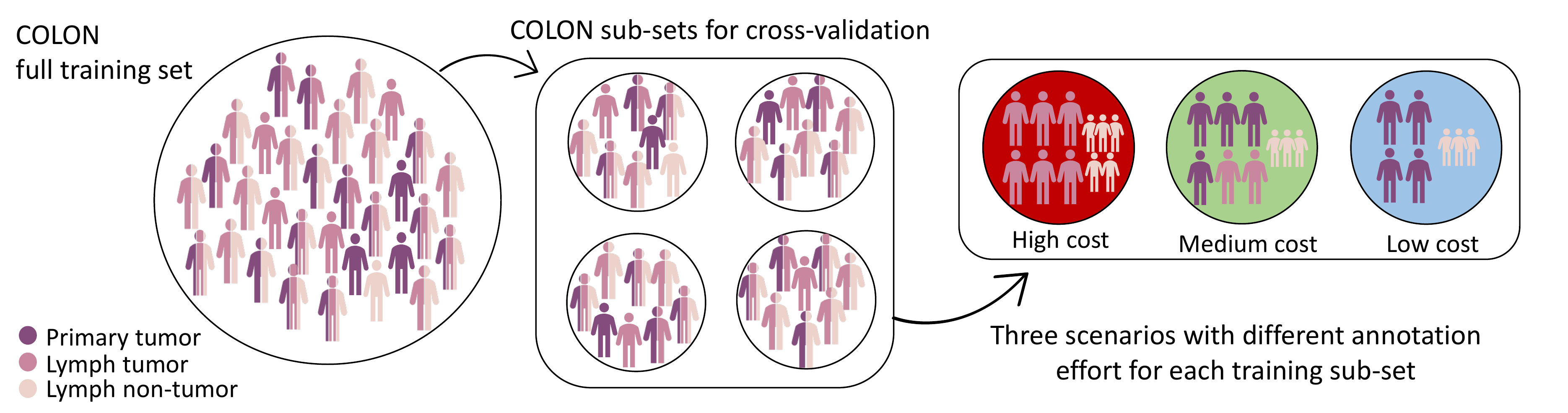}
    \caption{Colon training set and annotation cost scenarios overview. The full colon training set consists of all tissue types (left). Cross-validation of limited data access simulation is possible by dividing the full dataset into four sub-sets (middle). For each sub-set, three scenarios for annotation effort are created (right): high (only lymph node tissue), medium (primary and very little lymph node), and low (primary tumor).}
    \label{fig:lnco_subsets_experiments}
\end{figure}

To provide a deeper understanding of the impact of inter- and intra-organ augmentation strategies, we set up an experimental framework that evaluates different perspectives in terms of data availability and augmentation protocols. As illustrated in Figure~\ref{fig:data_exp_overview}, we propose to leverage the readiness of the primary colon cancer tumor, as well as already existing carcinoma datasets for different organs tissue (breast and skin). We evaluate different training data compositions for three different data availability scenarios of the target domain (colon lymph node metastasis), as illustrated in Figure~\ref{fig:lnco_subsets_experiments}. In what follows we outline the datasets, target scenarios, augmentation techniques, as well as evaluation protocol. For details on the experimental setup, we refer to the supplementary material.

\paragraph{\textbf{Datasets}}
In the conducted experiments, the target colon adenocarcinoma dataset consists of data from 37 anonymized patients, where data from 5 patients were used as the test set, and the rest used for training and validation~\cite{Maras_2019}. The dataset contains images from both primary and lymph node tumor samples, as well as lymph node non-tumor tissue.
For the inter-organ augmentations, we selected breast and skin carcinoma, driven by their high clinical occurrence, existing datasets availability, and cancer type/similarity compared to colon adenocarcinoma. The breast dataset~\cite{litjens_2018a} consists of whole slide images of sentinel breast lymph node sections, containing breast cancer metastasis. This cancer type, originating from epithelial cells, is similar to colon cancer. 
On the other hand, the skin cancer dataset~\cite{Lindman_2019,Stadler_2021} consists of abnormal findings identified as basal cell carcinoma, squamous cell carcinoma, and squamous cell carcinoma in situ. These tissue and cancer types are more different from regional colon lymph nodes and colon adenocarcinoma. The whole-slide images of all three datasets were sampled to extract patches. The data were extracted at a resolution of $0.5$ microns per pixel, with a size of $256\times256$. All three datasets are publicly available for use in legal and ethical medical diagnostics research.

\paragraph{\textbf{Target scenarios}}
To simulate limited access of target domain training data, but also cross-validate the experiments' performance and the outcome observations, the available full colon dataset was divided into four subsets, ensuring balance between the different tissue types, as well as number of patients. 
Each sub-set has non-tumor lymph node tissue data from at least five patients, tumor lymph node samples from at least six patients, and primary tumor samples from at least four patients. Considering the different costs of annotation effort of the primary tumor and lymph node tissue we identify three baseline experiments: 1) the \textbf{high} cost scenario including only lymph node tissue data, 2) the \textbf{medium} cost including primary tumor data along with lymph node tissue from only two patients on average, and 3) the \textbf{low} cost case including only primary tumor (i.e., no target tumor representation) and lymph node non-tumor tissue from just two patients on average (Figure~\ref{fig:lnco_subsets_experiments}). The inter-organ augmentations do not charge the baseline experiments with extra annotation effort for the experts, since they utilize already available annotated data. 

\paragraph{\textbf{Augmentation strategies}}
In order to augment the target dataset with intra- and inter-organ data, we consider two different strategies:
1) direct mix of source and target training data, and 2) image synthesis where the augmented samples are adaptations from one organ's data distribution to the target domain through a Cycle-GAN image-to-image-translation~\cite{Zhu_2017}. Furthermore, to evaluate the optimal ratio between augmented and target data, we investigate augmenting the dataset with either equal amount (i.e., doubling the total training set size), or half the amount. 

While the direct mix allows for understanding of how data from a different organ impact the target domain, the domain-adaptation of images demonstrates if there are features in the source domain that can be utilized if the data distribution is aligned with the target domain. Although there are other strategies for aligning the domains, such as stain normalization~\cite{macenko_2009}, the Cycle-GAN provides us with a representative method for investigating the performance of transformed source data.
Note that since the target domain is formulated in three different scenarios, the different inter- and intra-organ augmentations are evaluated in three separate experiments, i.e., Cycle-GANs need to be trained separately for each target scenario.

\paragraph{\textbf{Evaluation protocol}}
To evaluate task performance for the different combinations of training data, we train a deep classifier for tumor detection and evaluate it on the lymph node colon adenocarcinoma metastasis test data. The network consists of three convolutional and two fully connected layers with dropout and batch-normalization for regularization. We employ standard geometric and color jittering augmentations. The networks are trained with Adam optimizer 
for 50 epochs, out of which the best model is selected. 
For the GAN augmentation, we used a vanilla Cycle-GAN\footnote{\url{https://github.com/vanhuyz/CycleGAN-TensorFlow}}, trained for $250,000$ iterations, using colon data defined per experiment and the organ's entire dataset.  

We cross-validate each target scenario between the four colon dataset sub-sets. Moreover, we run each experiment's convolutional network five times to ensure adequate statistical variation. This work puts special emphasis on formulating augmentation schemes that lead to stable and robust results, highlighting that they are at least equally important as factors like training data size and downstream task reported performance.

In addition to the evaluation in terms of classification performance we also include a measure of the representation shift between source and target data~\cite{Stacke_2019,Stacke_2021}. This measure takes the distributions of layer activations over a dataset in a classifier, and compares this between the source and target domains, capturing the model-perceived similarity between the datasets. 

%% file: results.tex
\section{Results}
\begin{figure}[t!]
    \centering
    \includegraphics[width=.45\textwidth]{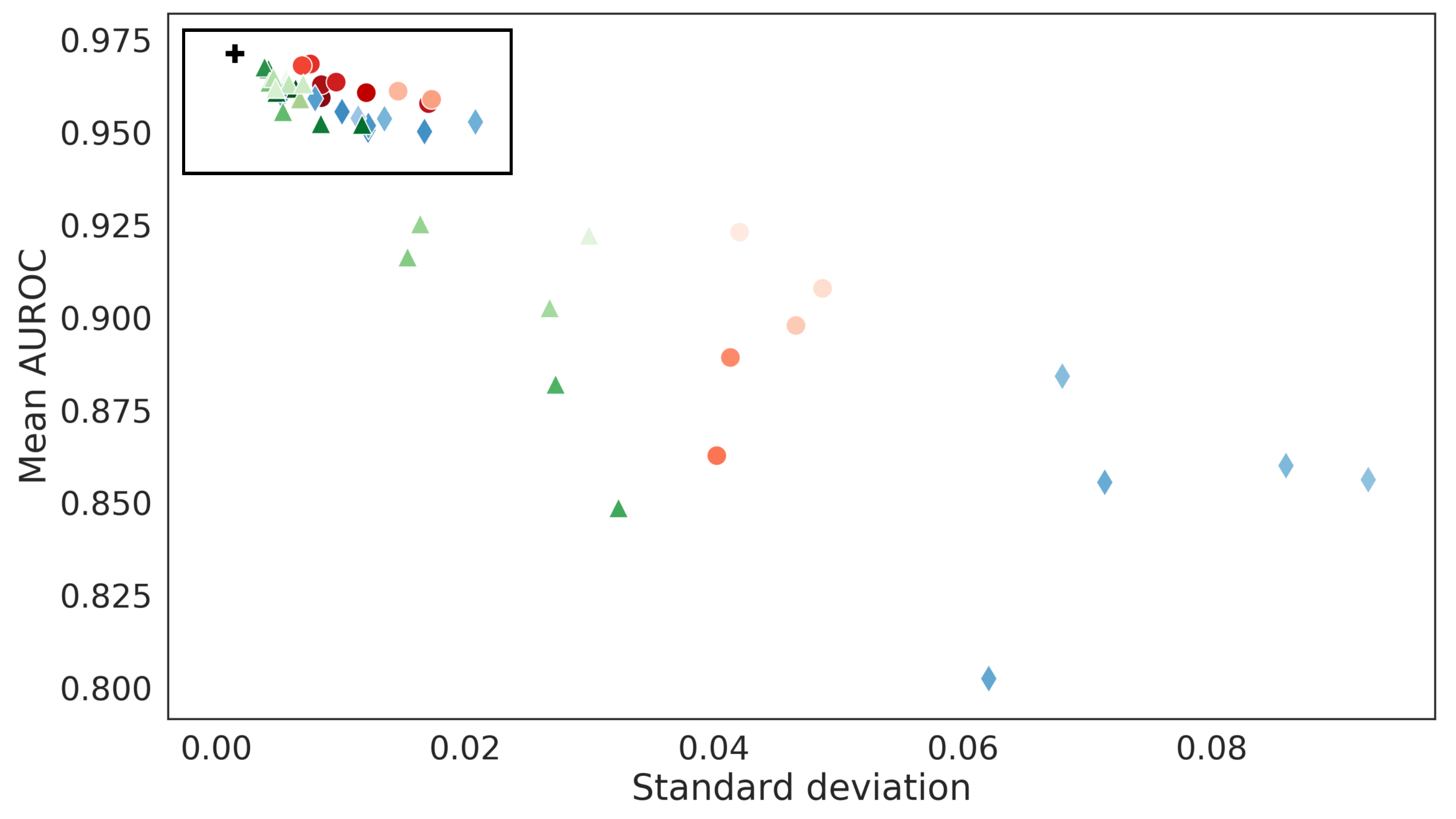}
    \includegraphics[width=.35\textwidth]{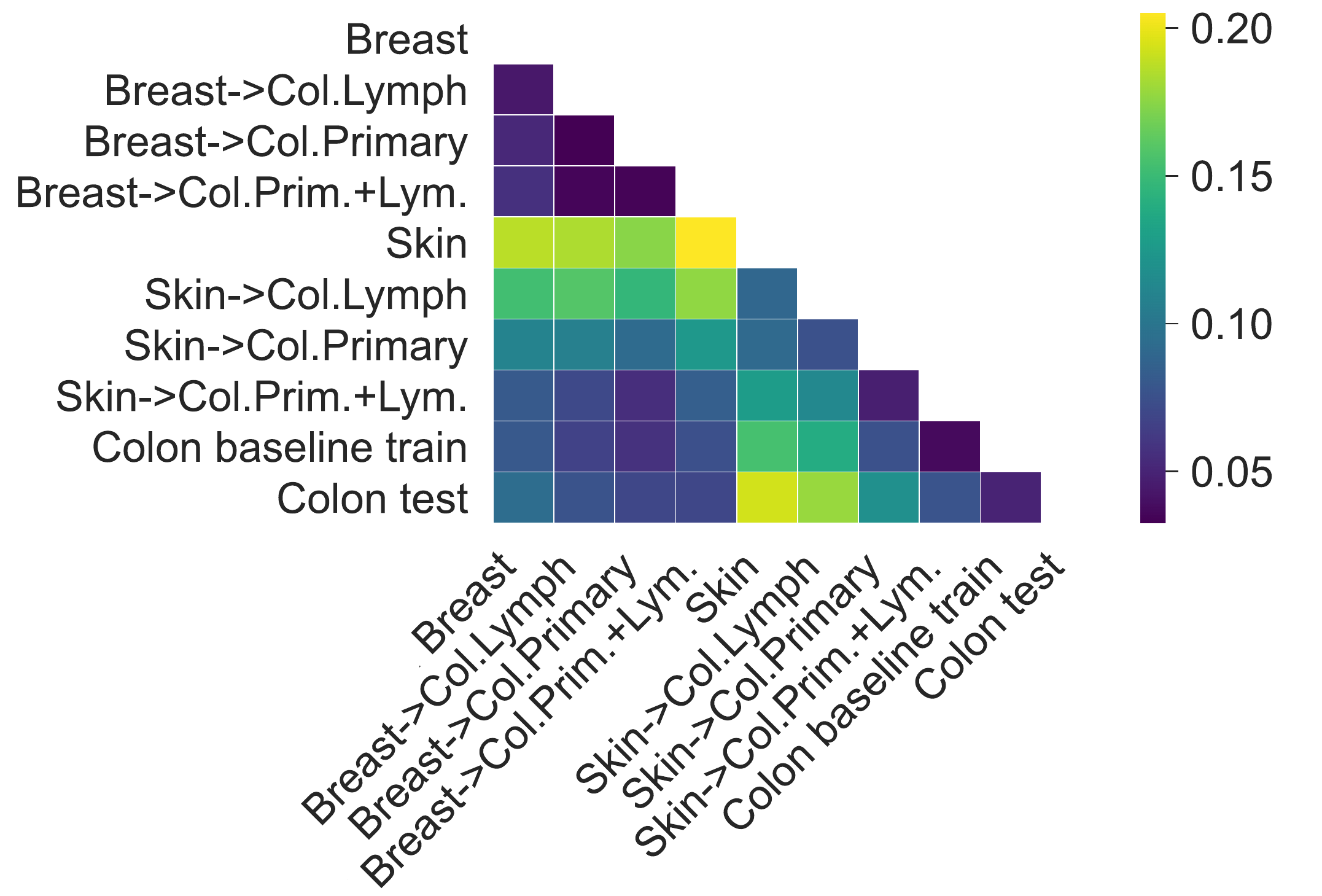}\\
    \includegraphics[width=.9\textwidth]{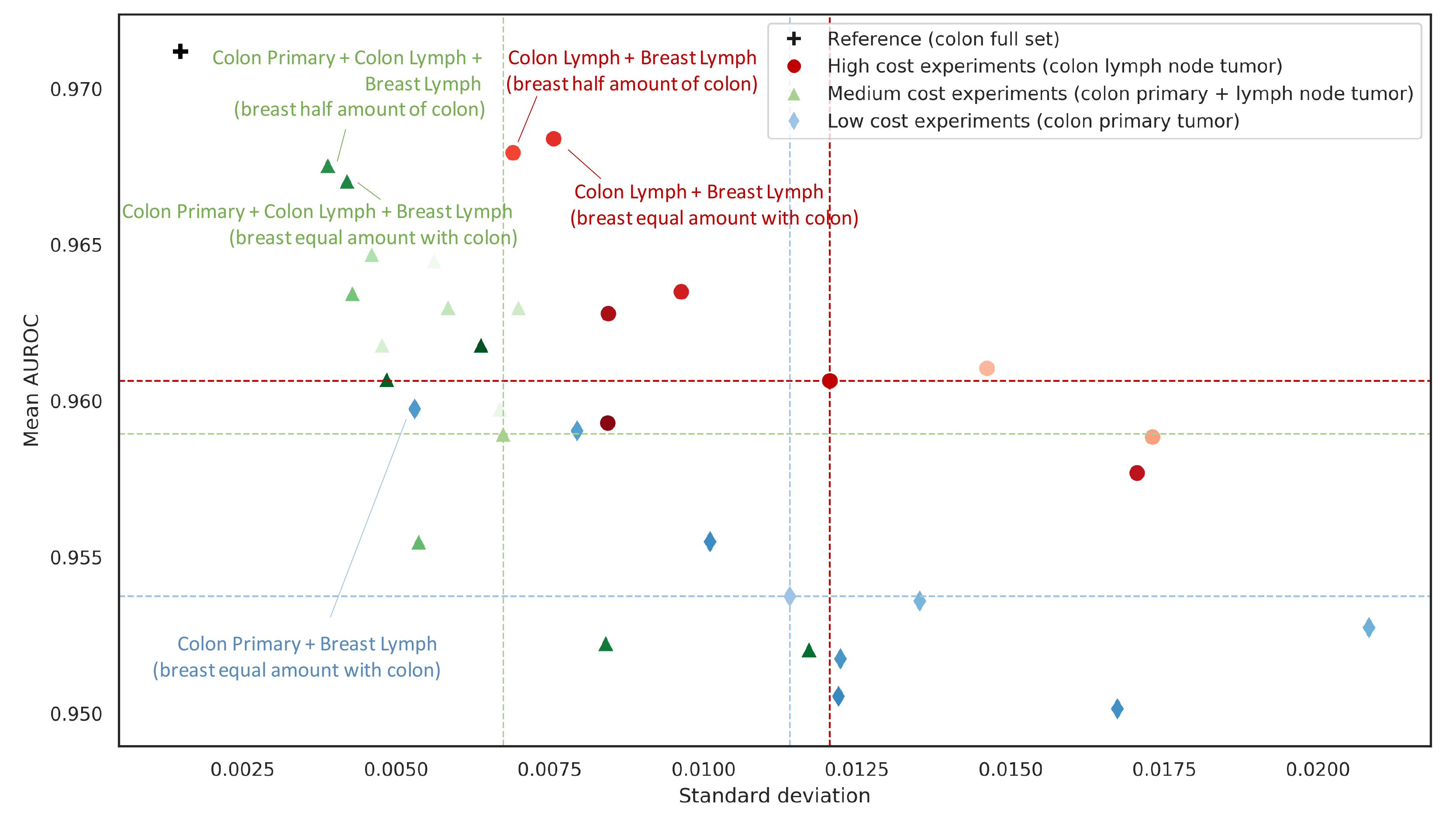}
    \caption{Mean AUROC to standard deviation for all the experiments (top left), and detailed view for the best performing ones with respect to the corresponding baselines (bottom). The representation shift (top right) is measured between different datasets using a classifier trained for the baseline medium effort scenario.}
    \label{fig:mean_stddev}
\end{figure}

\begin{table}[b!]
    \centering
    \caption{Mean AUROC for the best performing augmentation strategies for all three annotation effort scenarios.  Bre./Skin$\rightarrow$Col. and Prim.$\rightarrow$Lym. denote data domain transformation using Cycle-GAN,
    and (equal/half am.) is the amount of added images in relation to the size of the available baseline training set.}
    \label{tab:best_experiments_mean_auroc_stddev}
    \begin{adjustbox}{max width=.95\textwidth}
        \begin{tabular}{rccc}
         & \multicolumn{3}{c}{Mean AUROC$\pm$ stddev}\\ \cmidrule(lr){2-4}
        Augmentation & High & Medium & Low \\
        \toprule
         & 0.9607$\pm$0.01206 &0.9589$\pm$0.0067 & 0.9538$\pm$0.0114 \\ \cmidrule(lr){1-4}
        +Breast(equal am.) & 0.9684$\pm$0.0076 & 0.9671$\pm$0.0042 & 0.9598$\pm$0.0053 \\
        +Breast(half am.) & 0.9680$\pm$0.0069  & 0.9676$\pm$0.0039 & 0.9591	$\pm$0.0079 \\
        +Bre.$\rightarrow$Col.(equal am.) & 0.9577$\pm$0.0171 & 0.9521$\pm$0.0117 & 0.9502$\pm$0.0167 \\
        +Bre.$\rightarrow$Col.(half am.) & 0.9635$\pm$0.0096  & 0.9523$\pm$0.0084 & 0.9518$\pm$0.0122 \\
        +Skin$\rightarrow$Col.(equal am.) & 0.9589$\pm$0.0173 & 0.9555$\pm$0.0054 & 0.9528$\pm$0.0208 \\
        +Skin$\rightarrow$Col.(half am.) & 0.9611$\pm$0.0146 & 0.9635$\pm$0.0043 & 0.9536$\pm$0.0135 \\
        +Prim.$\rightarrow$Lym. & -- & 0.9630$\pm$0.0070 & --
        \end{tabular}
    \end{adjustbox}
\end{table}

The experimental setup with different scenarios, data, augmentation strategies, and amount of augmented data, lead to a total of 50 individual experiments. 
Figure~\ref{fig:mean_stddev} plots the mean AUROC (Area Under the Receiver Operating Characteristic) curve against the standard deviation, computed over the 4 sub-sets' performance for 5 trainings per sub-set, for the different cost scenarios, as well as the representation shift between different datasets.  
Baselines are noted with dashed line crosses and the different augmentation experiments for each scenario are color and marker coded. 
The best performing setups are reported in Table~\ref{tab:best_experiments_mean_auroc_stddev}, where each column corresponds to a different scenario (marked with different colors in Figure~\ref{fig:mean_stddev} top left and bottom). For an exhaustive presentation of the numbers and legends for all the experiments, we refer to the supplementary material. In what follows, we will focus on the most interesting observations made from the results in Figure~\ref{fig:mean_stddev} and Table~\ref{tab:best_experiments_mean_auroc_stddev}.

\subsection{Primary tumor is an inexpensive training data source for lymph node colon cancer metastasis detection}
Inspecting the baseline performances of the three scenarios in Figure~\ref{fig:mean_stddev} and Table~\ref{tab:best_experiments_mean_auroc_stddev}, it is clear that the most robust approach is the medium annotation effort scenario, despite having target domain representation by only $1/3$ of the training set, originating from a small number of patients. Compared to the high-cost baseline, the medium effort offers significantly improved robustness (45\% decrease in standard deviation) while maintaining the same AUROC performance (0.2\% drop). 
Moreover, the low-cost baseline augmented with breast lymph node tissue achieves even better stability ($21\%$ further decrease in standard deviation) for the same AUROC ($0.1\%$ drop over the high cost). This is supported by the various medium-cost augmentations that exceed the performance of all baselines, and prove to be the most cost-effective among all the tested experiments (Figure~\ref{fig:mean_stddev} bottom). This shows that utilizing the primary tumor data, even with no target domain representation, to detect lymph node metastasis of colon adenocarcinoma is possible. This paves the way for similar possibilities in other cancer types using the TNM staging protocol \cite{brierley_2016}, such as breast~\cite{fitzgibbons_2000} and esophagus~\cite{ajani_2019}.

\subsection{Domain adapted data closes the gap for large representation shift between source and target domain}
By inspecting the confusion matrix of representation shifts between datasets and the performance of the augmentation strategies in Figure~\ref{fig:mean_stddev}, we observe that image-to-image translation of tissues with already low representation shift compared to the target distribution (e.g., breast tissue), does not improve performance or robustness. 
In this case, direct mix augmentation increases the data diversity, sufficiently to outperform the baselines. 
Skin data on the other hand, which exhibits a much larger representation shift from the colon lymph node metastasis samples, show a significant avail from data adaptation. Direct mix augmentation with such a dataset drastically decreases performance, indicating that the added diversity does not contribute to convergence. Domain adaptation closes the gap in the representation shift, leading improved performance as compared to the baseline. 

\subsection{Robustness is improved by out-of-domain data}
One of the central observations from the experiments regards the differences in robustness for different augmentation scenarios, with the robustness measured from the standard deviation over multiple trainings.
Classical image augmentation is the most critical component for increasing both generalization performance and robustness, and is applied in all of our different experiments.
In addition to this, both primary tumor data, as well as inter-organ augmentations using breast and skin data, can provide an additional boost in terms of performance. 
However, analyzing the relations between AUROC and standard deviation in Figure~\ref{fig:mean_stddev}, we can see a more pronounced impact on robustness.

As discussed above, the medium-effort scenario is on pair with the high-effort scenario in terms of AUROC, and gives overall lower variance. For the inter-organ augmentations, breast data do not benefit from adaptation by means of the Cycle-GAN, while this is essential for reaping the benefits of the skin data. These results point to how the out-of-domain data (e.g., primary tumor, or other organ tissue) can have a regularizing effect on the optimization, which has a significant impact on robustness. This means that in certain situations it is better not to perform data adaptation since this will decrease the regularizing effect (e.g., primary tumor data, or breast data with low representation shift). However, if the data is widely different (e.g., skin data, with large representation shift), it is necessary to perform adaptation in order to benefit from augmentation.

%% file: conclusion.tex
\section{Conclusion}

This paper presented a systematic study on the impact of inter- and intra-organ augmentations under different training data availability scenarios for lymph node colon adenocarcinoma metastasis classification. The results show that accuracy can be boosted by utilizing data from different organs, or from the primary tumor, but most importantly how this has an overall positive effect on the robustness of a model trained on the combined dataset.

One of the important aspects when incorporating data from a different domain is the strategy used for performing augmentation. For a source dataset that more closely resembles the target data, adaptation of the image content can have a detrimental effect, while for different data image adaptation is a necessity. For future work, it would be of interest to closer define when to adapt and when not to. This could potentially be quantified with the help of measures that aim at comparing model-specific differences between datasets, such as the representation shift used in these experiments. Moreover, there are other types of data and augmentation strategies that could be explored, as well as model-specific domain adaptation and transfer-learning techniques. We believe that utilization of inter-organ data formulations will be an important tool in future machine learning-based medical diagnostics.

\subsubsection{Acknowledgments}
We would like to thank Martin Lindvall for the interesting discussions and insights into the use of cancer type-specific primary tumor data for lymph node metastasis detection, and Panagiotis Tsirikoglou for the suggestions in results analysis. \\
This work was partially supported by the Wallenberg AI, Autonomous Systems and Software Program (WASP) funded by the Knut and Alice Wallenberg Foundation, the strategic research environment ELLIIT, and the VINNOVA grant 2017-02447 for the Analytic Imaging Diagnostics Arena (AIDA).   